\documentclass[twocolumn,showpacs,preprintnumbers,amsmath,amssymb,prc]{revtex4}
\usepackage{graphicx}   
\graphicspath{{./}{./figs/}}
\begin{document}
%
%
\title{New measurement of the d(d,p)t reaction at astrophysical energies via the Trojan-horse method}
%
%
\author{Chengbo Li}
\thanks{Supported by Beijing Natural Science Foundation (1122017) and National Natural Science Foundation of China (11075218, 10575132) }
\email{lichengbo2008@163.com}
\affiliation{Beijing Radiation Center,  Beijing 100875, China}
\affiliation{Key Laboratory of Beam Technology and Material Modification of Ministry of Education,College of Nuclear Science and Technology, Beijing Normal University, Beijing 100875, China}

\author{Qungang Wen}
\affiliation{Anhui University,  Hefei 230601, China}

\author{Yuanyong Fu}
\author{Jing Zhou}
\author{Shuhua Zhou}
\author{Qiuying Meng}
\affiliation{China Institute of Atomic Energy, Beijing 102413, China}

 \author{C. Spitaleri}
 \author{A. Tumino}
 \author{R. G. Pizzone}
 \author{L. Lamia}
 \affiliation{Laboratori Nazionali del Sud-INFN, Catania, Italy}
 \affiliation{Dipartimento di Fisica e Astronomia, Università degli Studi di Catania, Catania, Italy}
 \affiliation{Facoltà di Ingegneria e Architettura, Università degli Studi di Enna "Kore", Enna, Italy}

%

%
\date{\today}
\begin{abstract}
The study of d(d,p)t reaction is very important for the nucleosynthesis in both standard Big Bang and stellar evolution, as well as for the future fusion reactors planning of energy production.
The d(d,p)t bare nucleus astrophysical S(E) factor has been measured indirectly at energies from about 400 keV down to several keV by means of the Trojan horse method 
applied to the quasi-free process  $\rm {}^2H({}^6Li,pt){}^4He$ induced at a lithium beam energy of  9.5 MeV, which is closer to the zero quasi-free energy point. 
An accurate analysis leads to the determination of the $\rm S_{bare}(0)=56.7 \pm 2.0 keV \cdot b$ and of the corresponding electron screening potential  $\rm U_e = 13.2 \pm 4.3 eV$.
In addition, this work gives an updated test for the Trojan horse nucleus invariance comparing with previous indirect investigations using $\rm {}^3He=(d+p)$ breakup. 

\end{abstract}
%
\maketitle
\section{Introduction}

The d+d nuclear reactions are important in both nuclear astrophysics \cite{sbbn1984, sbbn1998,sbbnthmdd14} and fusion energy applications \cite{fus1983, fus1992}.

These reactions are among the thermonuclear processes occurring during the first minutes of the universe immediately after the Big Bang.
In particular, knowledge and modelling of the primordial abundance of deuterium, which depends on precise cross section data, give important information about the baryon density of the universe. Moreover, primordial deuterium is burned during the earliest evolution stage of stars: the pre-main sequence phase. Thus, a better knowledge of the parameters characterizing these reactions can improve our understanding of the first phases of stellar evolution.
As for the Standard Big Bang Nucleosynthesis, the region of interest ranges from 50 to 300 keV, and it is only several to 20 keV for stellar evolution processes.

In addition to these important astrophysical topics, the interest of scientists around reactions involving deuterium has been also triggered by the promising possibility of exploit them as a powerful and low-polluting source of energy in fusion reactors. 
These reactions belong to the network of processes inside the fusion reactors. 
These reactors are expected to operate in the temperature range of kT = 1-30 keV.

Several experiments have been performed below 200 keV, but available data are not always in agreement within each other and some of them are affected by large systematic errors. 
Another weak point is that available data below 10 keV, region of interest for fusion reactors as well as for burning deuteron in the Pre-Main Sequence phase of stellar evolution, are affected by the “electron screening”. 

For these reasons,  new indirect experimental studies were called for to provide new data in the full range of interest for pure and applied physics.
The Trojan Horse Method (THM) \cite{ gbaur1986, stypel2003} has been applied to the indirect study of the d+d reactions using $\rm ^3He=(d+p)$  and $\rm ^6Li=(d+\alpha)$ breakup \cite{thmdd2013,thmdd2014}, but the $\rm ^6Li$ breakup data give much less points and  larger errors than that in the case of $\rm ^3He$  breakup.

In this paper, we report on a new investigation of the d(d,p)t  reaction by means of the THM applied to the $\rm {}^2H({}^6Li,pt){}^4He$ quasi free process with a beam energy of 9.5 MeV, which is closer to the zero quasi-free energy point. 

\section{Trojan Horse Method}
The Coulomb barrier and electron screening cause difficulties in directly measuring nuclear reaction cross sections of charged particles at astrophysical energies. 
To overcome these difficulties, the THM \cite{ gbaur1986, stypel2003} has been introduced as a powerful indirect tool in experimental nuclear astrophysics \cite{ cspitaleri2003,  thmall2013,thmrev2011,thmrev2014,thmdist2009,thmplb1,thmplb2, thmdd2013,thmdd2014, cheng2005, sromano2006, wen2008, wen2011,thmg4}. 
The THM provides a valid alternative approach to measure unscreened low-energy cross sections of charged particle reactions. It can also be used to retrieve information on the electron screening potential when ultra-low energy direct measurements are available.

The basic assumptions of the THM theory have been discussed extensively in \cite{ stypel2003, cspitaleri2003,  thmall2013,thmrev2011,thmrev2014},  and the detailed theoretical derivation of the formalism employed can be found in \cite{stypel2003} . 

\begin{figure}
\begin{center}
\includegraphics[width = 0.35\textwidth]{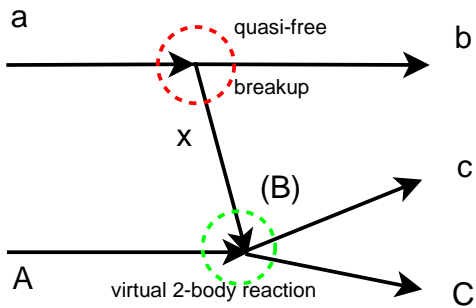}\\
\caption{(Color online) The schematic representation of Trojan-horse method}
\label{fig1} 
\end{center}
\end{figure}

A schematic representation of the process underlying the THM is shown in Figure \ref{fig1}.
The method is based on the quasi-free (QF) reaction mechanism, which allows one to derive indirectly the cross section of a two-body reaction Eq.(\ref{eq:two_body})
\begin{equation}\label{eq:two_body}
    A+x \rightarrow C+c
\end{equation}
from the measurement of a suitable three-body process under the quasi-free kinematic conditions: 
\begin{equation}\label{eq:three_body}
    A+a \rightarrow C+c+b
\end{equation}
where the nucleus $a$ is considered to be dominantly composed of clusters $x$ and $b$ ( $a=(x \oplus b) $).

After the breakup of nucleus $a$ due to the interaction with nucleus $A$, the two-body reaction  (Eq.(\ref{eq:two_body})) occurs only between nucleus $A$ and the transferred particle $x$ whereas the other cluster $b$ behaves as a spectator to the virtual  two-body reaction during the quasi-free process. 
The energy in the entrance channel $\rm E_{Aa}$ is chosen above the height of the Coulomb barrier $\rm E_{Aa}^{C.B.}$ , so as to avoid the reduction in cross section. 

At the same time, the effective energy $\rm E_{Ax}$ of the reaction between $A$ and $x$ can be relatively small, mainly because the energy $\rm E_{Aa}$ is partially used to overcome the binding energy $\varepsilon_a$ of $x$ inside $a$, even if particle $x$ is almost at rest the extra-energy is compensated by the binding energy of $a$  (Eq.(\ref{eq:Eqf1})), and the Fermi motion of $x$ inside $a$, $\rm E_{xb}$, is used to span the region of interest around $\rm E_{Ax}^{qf}$:
\begin{equation}\label{eq:Eqf1}
        E_{Ax}^{qf}=E_{Aa}\left(1-\frac{\mu_{Aa}}{\mu_{Bb}}\frac{\mu_{bx}^{2}}{m_x^2}\right)-\varepsilon_{a}
\end{equation}
\begin{equation}\label{eq:Eqf2}
        E_{Ax}=E_{Ax}^{qf}\pm E_{xb}
\end{equation}

Since the transferred particle $x$ is hidden inside the nucleus $a$ (so called Trojan-horse nucleus), it can be brought into the nuclear interaction region to induce the two-body reaction $A+x$, which is free of Coulomb suppression and, at the same time, not affected by electron screening effects.

Thus, the two-body cross section of interesting can be extracted from the measured quasi free three-body reaction inverting the following relation:
\begin{equation}\label{eq:sec3all}
\frac{d^3\sigma}{dE_{Cc}d\Omega_{Bb}d\Omega_{Cc}} = KF \cdot |W|^2  \cdot  \frac{d\sigma}{d\Omega}^{TH} 
\end{equation}
where KF is the kinematical factor,  $\rm |W|^2$ is the momentum distribution of the spectator $b$ inside the Trojan-horse nuclei $a$,   
and $\rm {d\sigma}/{d\Omega}^{TH}$ is the half-off-energy-shell (HOES) cross section of the two-body reaction $\rm A+x\rightarrow C+C$:
\begin{equation}\label{eq:sec2all}
\frac{d\sigma}{d\Omega}^{TH} = \sum_l C_l \cdot P_l  \cdot \frac{d\sigma_l}{d\Omega}({Ax\rightarrow Cc)}
\end{equation}
where $\rm \frac{d\sigma_l}{d\Omega}({Ax\rightarrow Cc)}$ is the real on-energy-shell cross section of the two-body reaction $\rm A+x\rightarrow C+c$ for the $l$ partial wave,  $\rm P_l$ is the penetration function caused by the Coulomb wave function, and $\rm C_l$ is the scaling factor.

\section{Experiment}

\begin{figure}
\begin{center}
\includegraphics[width = 0.35\textwidth]{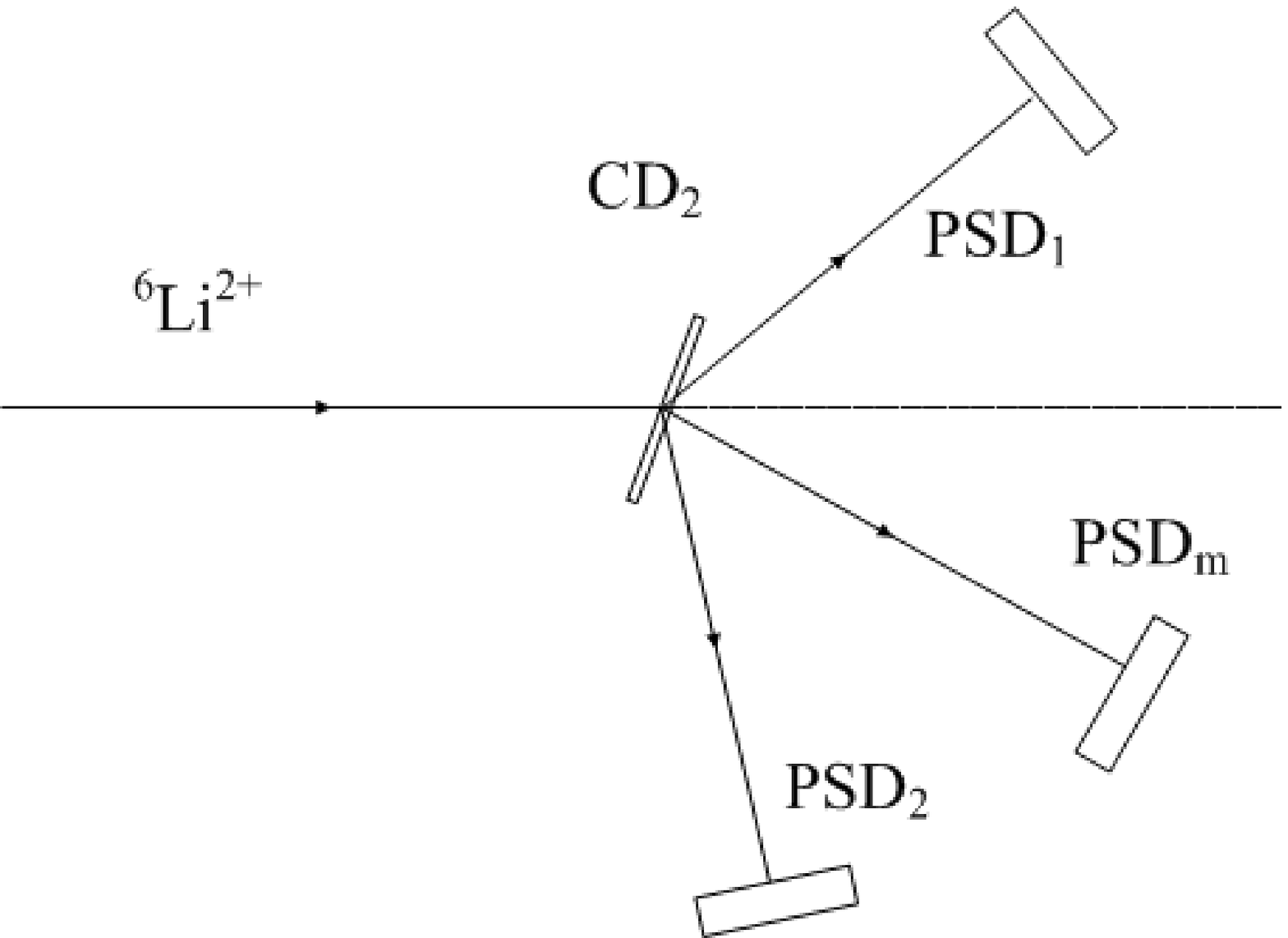}\\
\caption{Experiment setup of the $\rm {}^2H({}^6Li,pt){}^4He$ reaction }
\label{fig2} 
\end{center}
\end{figure}

The measurement of the $\rm {}^2H({}^6Li,pt){}^4He$ reaction was performed at the Beijing National Tandem Accelerator Laboratory at China Institute of Atomic Energy. 
The experimental setup was installed in the nuclear reaction chamber at the R60 beam line terminal as shown in Figure \ref{fig2}. 
The $\rm ^{6}Li^{2+}$ beam at 9.5 MeV provided by the HI-13 tandem accelerator was used to bombard a deuterated polyethylene target $\rm CD_{2}$. 
The thickness of the target is about $\rm 160 \mu g/cm^{2}$. In order to reduce the angle uncertainty coming from the large beam spot, a linear target with 1 mm width was used.

A position sensitive detector $\rm PSD_1$ was placed at $40^{\circ} \pm 5^{\circ}$ to the beam line direction and about 238 mm from the target to detect the outgoing particle triton ($t$), 
and another detector $\rm PSD_2$ was used at $78^{\circ} \pm 5^{\circ}$  in the other side of the beam line at 245 mm  from the target to detect the outgoing particle proton ($p$).
The arrangement of the experimental setup was modelled in Monte Carlo simulation in order to cover a region of quasi-free angle pairs.
A $\rm PSD_m$ was placed at $32^{\circ} \pm 5^{\circ}$ opposite to $\rm PSD_1$ as a monitor.
The energy resolution of the PSDs is about 0.6\%-0.8\% for 5.48 MeV $\alpha$ source. 

It should be mentioned that no $\rm \Delta E$ detector was mounted before PSDs, that was, no particle identification was performed in the experiment. 
It was beneficial to improve the energy and angular resolution without $\rm \Delta E$ detectors, because it would lead to additional straggling of energies and angles when particles passing through these detectors. 
While it was easy to select events for this reaction without particle identification from the kinematics in data analysis with the help of simulation. 

The trigger for the event acquisition was given by coincidence of signals by $\rm Gate=PSD_1\times(PSD_2+PSD_m)$.
Energy and position signals for the detected particles were processed by standard electronics and sent to the acquisition system MIDAS for on-line monitoring and data storage for off-line analysis.
In order to perform position calibration, a grid with a number of equally spaced slits was placed in front of each PSD for calibration runs.

\section{Data analysis and results}

The position and energy calibration of the detectors  were performed using elastic scatterings on different targets ($\rm ^{197}Au$, $\rm ^{12}C$, and $\rm CD_2$) induced by a proton beam at energies of 6, 7, 8 MeV.  A standard $\alpha$ source of 5.48 MeV was also used. 

After the calibration of the detectors, the energy and momentum of the third  undetected particle ($\alpha$)  were reconstructed from  the complete kinematics of the three-body reaction $\rm ^6Li + d \rightarrow t+p+ \alpha$, under the assumption that the first particle is a triton (detected by $\rm PSD_1$)  and the second one is a proton (detected by $\rm PSD_2$).

\subsection{Selection of the three body reaction events}

\begin{figure}
\begin{center}
\includegraphics[width = 0.40\textwidth]{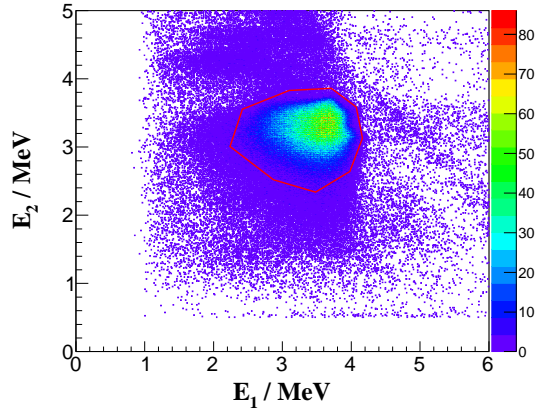}\\
\caption{(Color online) Selection of the three body reaction events of $\rm {}^2H({}^6Li,pt){}^4He$ from the $\rm E_1- E_2$ kinematic locus}
\label{fig3} 
\end{center}
\end{figure}

The basic step of data analysis is to select the three-body reaction events of $\rm {}^2H({}^6Li,pt){}^4He$ from all exit channels.
Figure \ref{fig3}  shows the experimental spectrum of the $\rm E_1- E_2$ kinematic locus. Comparing with the Monte Carlo simulation \cite{thmg4}, we can select the range by a graphical cut (red line polygon) where the three body reaction $\rm {}^2H({}^6Li,pt){}^4He$ events are located.
It will be used as a basic selection cut in the following data analysis.

\subsection{$\rm Q_3$ value spectum}
\begin{figure}
\begin{center}
\includegraphics[width = 0.45\textwidth]{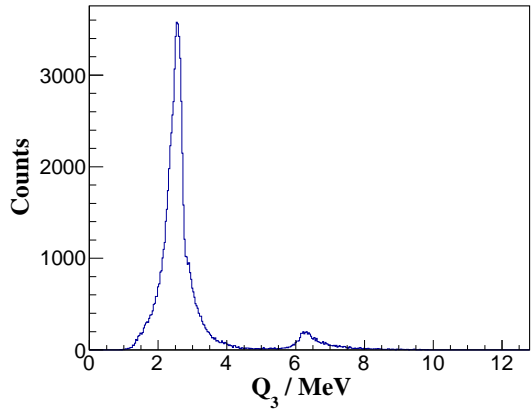}\\
\caption{(Color online) Experimental $\rm Q_3$ value spectum from the selection of Fig.\ref{fig3} for the kinematic locus of $\rm {}^2H({}^6Li,pt){}^4He$. The relevant peak is the one at about 2.5 MeV.}
\label{fig4} 
\end{center}
\end{figure}

Once selected the three-body reaction events of $\rm ^{2}H(^{6}Li, pt)^{4}He$, the experimental $\rm Q_3 $ value can be extracted, as reported in Figure \ref{fig4}. 

There is a peak whose centroid is at about 2.5 MeV (in good agreement with the theoretical prediction, Q = 2.558 MeV). 
It is a clear signature of the good calibration of detectors as well as of the correct identification of the reaction channel. 

The events outside of the 2.5 MeV peak came from the background and some reaction located in the energy range in Figure \ref{fig3} but not in agreement with the assumption that the first particle is a triton and the second one is a proton, so that the calculated $\rm Q_3$ value deviated from the expected value. 

Only events inside the 2.5 MeV Q-value peak were considered for the further analysis.

\subsection{Momentum distribution of $\alpha$ inside $\rm ^6Li$ }

As in all standard THM analysis, the next step is to identify and separate the quasifree mechanism from all the other processes.
This is usually done by recalling the definition of a QF reaction, i.e., a reaction where the third particle (spectator) retains the same momentum it had  within the Trojan horse nucleus. 
Thus, the momentum distribution of the third and undetected particle will be examined. 
This gives a major constraint for the presence of the quasifree mechanism and the possible application of the THM. 

In order to extract the experimental momentum distribution of the spectator in the system where the Trojan horse particle $b$ is at rest,  narrow energy and angular windows should be selected.
Since $\rm (d\sigma/d\Omega )^{TH}$ is nearly constant in a narrow energy and $\theta_{c.m.}$ window, 
one can obtain the shape of the momentum distribution $\rm |W|^2$ of the undetected particle directly from the three-body reaction yield divided by the kinematical factor $\rm KF$, according to Eq.(\ref{eq:sec3all}).

\begin{figure}
\begin{center}
\includegraphics[width = 0.45\textwidth]{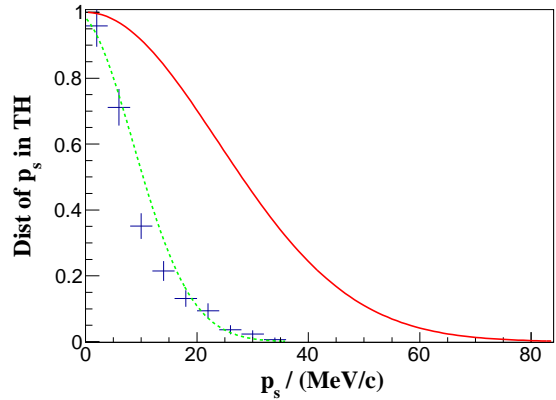}\\
\caption{(Color online) Experimental spectrum of momentum distribution for intercluster motion of $\alpha$ inside $\rm ^6Li$ (the blue points and the dotted green line for the fitting curve) compared with  the theoretical calculation (the red line)}
\label{fig5} 
\end{center}
\end{figure}

The obtained momentum distribution is reported in Figure \ref{fig5}, where it is compared with the theoretical prediction of the spectator momentum distribution, 
obtained using the Woods-Saxon potential with the standard geometrical parameters.

An evident distortion of the momentum distribution shows up and its measured full width at half maximum (FWHM) turns out to be around 23 MeV/c which is much smaller than the expected prediction of 72 MeV/c.  
This evidence was already observed in Ref. \cite{ thmdist2009}, where the width of the momentum distribution  for the spectator inside the Trojan horse nucleus was studied as a function of the transferred momentum $q_t$ from the projectile $a$ to the center of mass of the final system $\rm B = C + c$. 
In the present case, the value of $q_t$ is about 133 MeV/c, and the width of the momentum distribution is about 23 MeV/c. It is in agreement with the trend of  the curve that represents the best fit to the function reported in Ref.\cite{ thmdist2009}, $W_{FWHM}(q_t)=f_0 [1-\exp(-q_t/q_0)] $.

For the further analysis, the condition of $\rm |p_{s}|<20 MeV/c$ will be added to the above cuts to select the quasi-free events of the three-body reaction.

\subsection{S(E) factor and $\rm U_e$}

The last step is to extract the energy trend of the S(E) factor by means of the standard procedure of the THM after selecting the qusi-free three-body reaction events. 

Therefore, Eq.(\ref{eq:sec3all})  and Eq.(\ref{eq:sec2all}) are applied. 
The relevant two-body reaction cross section can be extracted from the measured three-body cross section with cut of selecting the quasi-free events from the three-body reaction.
Then, the S(E) factor can be determined from the definition of $\rm S(E)= \sigma (E) E \exp(2\pi\eta)$, where  the Sommerfeld parameter is $\rm \eta=Z_1 Z_2 e^2 / (\hbar v)$.
In present work, only the s-wave ($l$=0) was considered for the energy range of $\rm E_{cm}=0 - 400 keV$. 

\begin{figure} 
\begin{center}
\includegraphics[width = 0.33\textwidth,angle=-90]{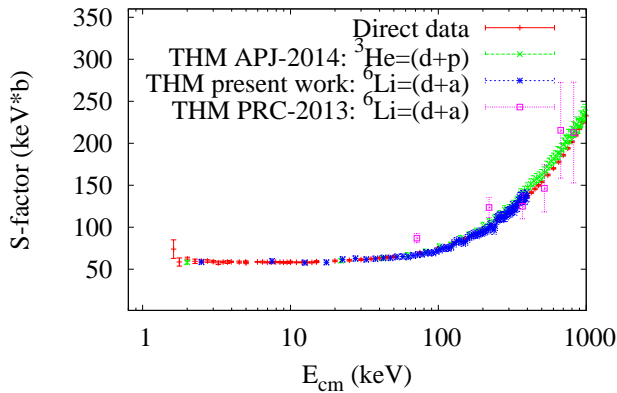}\\
\caption{(Color online) The S(E) factor obtained from THM measurement compared with direct data}
\label{fig6} 
\end{center}
\end{figure}

The results for the $\rm d(d, p)t$ reaction in terms of the bare nucleus astrophysical $\rm S_{bare}(E)$ factor are presented in Figure \ref{fig6} (blue points) after normalization with direct data (red points) \cite{thmdd2014, direct}. 
The normalization was performed in the energy range of $\rm E_{cm}=40 - 400 keV$, in which the electron screening effect is still neglectable.
It should be pointed out that direct data suffer from the electron screening effect which does not affect the THM results. Thus, the S(E) extracted via THM was called $\rm S_{bare}(E)$.
A polynomial fit was then performed on the data giving $\rm S_{bare}(0)=56.7 \pm 2.0 keV \cdot b$.

The data from the present experiment (blue points) are compared with those from PRC-2013 \cite{thmdd2013} of $\rm ^6Li=(d+\alpha)$ breakup in a previous THM experimental run (pink points) and those from APJ-2014 \cite{thmdd2014} of $\rm ^3He=(d+p)$ beakup experiment (green points).
An overall agreement is present among both direct and indirect data sets, within the experimental errors.

It should be pointed out that the errors in the present case are much smaller than in the case of  PRC-2013 \cite{thmdd2013}  using the same Trojan horse with higher beam energy and $\rm \Delta E$ detectors for particle identification.

It is also in agreement, within the experimental errors, with the result using a different Trojan horse  $\rm ^3He$  \cite{thmdd2014}.
That is, data extracted via the THM applied to  $\rm ^6Li$  and $\rm ^3He$ breakup are comparable among themselves. 
The Trojan horse particle invariance is confirmed in an additional and independent case  which was already observed in Ref. \cite{ths}.

The lack of screening effects in the THM $\rm S_{bare}(E)$ factors gives the possibility to return the screening potential $\rm U_e$ from comparison with direct data using the following screening function with $\rm U_e$ as free parameter. 
\begin{equation}\label{eq:Ue}
    f_{lab}(E)=\sigma_s(E)/\sigma_b(E)\simeq \exp(\pi \eta U_e /E)
\end{equation}

\begin{figure}
\begin{center}
\includegraphics[width = 0.33\textwidth,angle=-90]{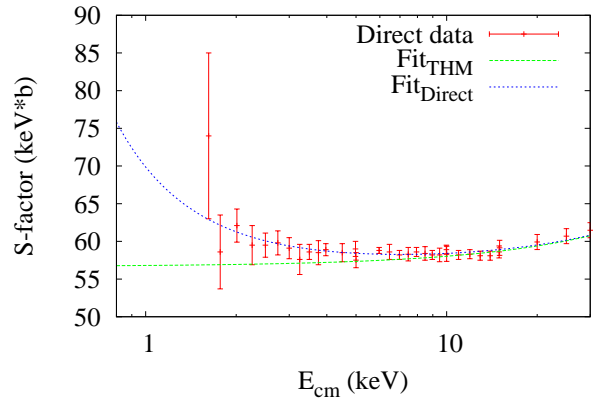}\\
\caption{(Color online) Fitting of S(E) to obtain $\rm U_e$}
\label{fig7} 
\end{center}
\end{figure}

The result is shown in Figure \ref{fig7}. The red points are the direct data by Greife et al. (1995) \cite{direct}. The blue dashed line is the fitting curve of direct data (screened), and the green line is the fitting of THM data (unscreened) of the present work. 
Thus, we obtain a value of $\rm U_e = 13.2 \pm 4.3 eV$, which is also in agreement with the one of Ref. \cite{thmdd2014} $\rm U_e = 13.4 \pm 0.6 eV$ . 

\section{Summary}

A new investigation of the $\rm ^2H({}^6Li,pt){}^4He$ reaction measurement to extrat information on the astrophysical $\rm S_{bare}(E)$ factor and screening potential $\rm U_e$ for the $\rm d(d, p)t$ reaction via the THM is presented in the present paper, shown in Table \ref{tab:THMdata}. 

An overall agreement within the experimental errors is present among both direct and indirect data sets using different Trojan horse nuclei.

The errors in the present case are much smaller than in the case of  PRC-2013 \cite{thmdd2013}, which uses the same Trojan horse $\rm ^6Li$.

In addition, the data extracted via the THM applied to  $\rm ^6Li$  and $\rm ^3He$ breakup are comparable among themselves. 
That is, the use of a different spectator particle does not influence the THM results.
Thus, this work  gives an updated test for the Trojan horse nucleus invariance. 

\begin{table}
    \caption{\label{tab:THMdata} Comparison of d(d,p)t indirect study via THM.}
    \begin{ruledtabular}
        \begin{tabular}{cccccc}
            Work & TH & $\rm E_0$ & $\rm E_{Ax}^{qf} $& $\rm S_0(E)$ & $\rm U_e $ \\
            &  &  (MeV) &  (MeV) &  ($\rm keV \cdot b$) &  (eV)\\
            \hline
             Present  & $\rm ^6Li=(d+\alpha)$  & 9.5  & 0.089 & $56.7 \pm 2.0$ & $13.2 \pm 4.3$\\
             \cite{thmdd2013} & $\rm ^6Li=(d+\alpha)$  & 14  & 0.866 & $75 \pm 21$ & -\\
            \cite{thmdd2014} & $\rm ^3He=(d+p)$  & 17  & 0.178 &  $57.7 \pm 1.8$ & $13.4 \pm 0.6$\\
      \end{tabular}
    \end{ruledtabular}
\end{table}

\begin{acknowledgments}
We thank Dr. Chengjian Lin and his research group, Dr. Xia Li, and the CIHENP research group in CIAE for their kind help during the experiment measurement. 
We also thank the staff of the HI-13 tandem accelerator laboratory for providing experimental beam and targets.
\end{acknowledgments}


\begin{thebibliography}{100}
\bibitem{sbbn1984} 
W. A. Fowler, Rev. Mod. Phys., \textbf {56}  (1984) 149

\bibitem{sbbn1998} 
D. N. Schramm and M. S. Turner, Rev. Mod. Phys. \textbf {70} (1998) 303.

\bibitem{sbbnthmdd14} 
R.G. Pizzone et al. , Ap. J., \textbf{786} (2014) 112

\bibitem{fus1983} 
R. E. Chrien, R. Kaita, J.D. Strachan, Nucl. Fusion, \textbf {23}, (1983) 1399

\bibitem{fus1992} 
H. S. Bosch, G. M. Hale,  Nucl. Fusion,  \textbf {32},  (1992) 611

\bibitem{gbaur1986} 
G. Baur, Phys. Lett. B \textbf {178} (1986) 135.

\bibitem{stypel2003} 
S.Typel, G.Baur. Annals Phys, \textbf{305} (2003)  228.

\bibitem{cspitaleri2003} 
C. Spitaleri, S. Cherubini, et al., Nucl.Phys. A, \textbf{719} (2003) 99c.

\bibitem{thmall2013} 
A. Tumino, C. Spitaleri, S. Cherubini, et al.,  Few-Body Syst, \textbf{54} (2013) 745.

\bibitem{thmrev2011} 
E. G. Adelberger, A. Garcia, R. G. Hamish Robertson, Rev. Mod. Phys., \textbf{83} (2011) 195.

\bibitem{thmrev2014} 
R E Tribble, C A Bertulani, M La Cognata,  Rep. Prog. Phys., \textbf{77}  (2014) 106901.

\bibitem{thmdist2009} 
R. G. Pizzone, C. Spitaleri, A. M. Mukhamedzhanov, et al., Phys. Rev. C,  \textbf{80} (2009) 025807

\bibitem{thmplb1} 
A. Tumino, C. Spitaleri, A.M. Mukhamedzhanov et al., Phys. Lett. B,   \textbf{700} (2011) 111 

\bibitem{thmplb2} 
A. Tumino, C. Spitaleri, A.M. Mukhamedzhanov et al., Phys. Lett. B,   \textbf{705} (2011) 746

\bibitem{thmdd2013} 
R. G. Pizzone, C. Spitaleri, C. A. Bertulani, et al., Phys. Rev. C,  \textbf{87} (2013)  025805 

\bibitem{thmdd2014} 
A. Tumino, R. Sparta, C. Spitaleri,  et al., The Astrophysical Journal, \textbf{785} (2014) 96.

\bibitem{cheng2005} 
Li Chengbo, R.G. Pizzone, C. Spitaleri, et al., Nuclear Physics Review, \textbf{22} (2005) 248.

\bibitem{sromano2006} 
S. Romano, L. Lamia, C. Spitaleri, et al., Eur. Phys. J. A, \textbf{27} (2006) 221.

\bibitem{wen2008} 
Qun-Gang Wen, Cheng-Bo Li, Shu-Hua Zhou,  et al., Phys. Rev. C,  \textbf{78} (2008) 035805.

\bibitem{wen2011} 
Qun-Gang Wen, Cheng-Bo Li, Shu-Hua Zhou, et al.,  J. Phys. G: Nucl. Part. Phys, \textbf{38} (2011) 085103.

\bibitem{thmg4} 
Cheng-bo Li, Qun-gang Wen, Shuhua Zhou, et al., Chinese Physics C,  \textbf{39} (2015) 054001.

\bibitem{direct} 
U. Greife,  F. Gorris,  M.  Junker, et al., Z. Phy. A, \textbf{351}  (1995) 107

\bibitem{ths} 
R. G. Pizzone et al., Phys. Rev. C,  \textbf{83} (2011) 045801

\end{thebibliography}
\end{document}